\documentclass[11pt]{article}
\usepackage{epsfig}
\title{{\bf Unveiling Dark Matter with HI and H$\alpha$ Data - \\
Observational Problems}}
\author{G.~Gentile$^{1}$, D. Vergani$^1$, P.~Salucci$^2$, P. Kalberla$^1$, U.~Klein$^1$ \\
\vspace{0.1cm}\\
\normalsize $^1$Radioastronomisches Institut Univ. Bonn, Auf dem H\"ugel 71, D-53121 Bonn, Germany\\
\normalsize $^2$SISSA, via Beirut 4, I-34013 Trieste, Italy \\
}
\date{}
\begin{document}
\maketitle

\def\bull{\vrule height .9ex width .8ex depth -.1ex}
\makeatletter
\def\ps@plain{\let\@mkboth\gobbletwo
\def\@oddhead{}\def\@oddfoot{\hfil\tiny
``ALMA Extragalactic and Cosmology Science Workshop on Dark Matter'';
Observatoire de Bordeaux, France, 22-24 May 2002}%
\def\@evenhead{}\let\@evenfoot\@oddfoot}
\makeatother

\begin{abstract}\noindent

We present combined H$\alpha$+HI rotation curves 
for a sample of spiral galaxies.
Most of the velocity profiles (spectra at single points)
in these galaxies are asymmetric, preventing the use of standard
methods like the first moment analysis and the single Gaussian fitting.
We thus propose a method similar to the Envelope-Tracing
method (Sofue \& Rubin 2001) to analyse those profiles 
from which we obtain HI rotation curves in good agreement
with the H$\alpha$ rotation curves. These final rotation curves provide the
required high resolution in the inner parts of the galaxies, but
also extend out to typically 2-3 R$_{\rm opt}$. They will hence allow us
to investigate the distribution of dark matter.
\end{abstract}

\section{Introduction}

The study of rotation curves of spiral galaxies provides the
best evidence for dark matter on galactic scales. However, important 
properties like the shape of the dark halo, the distribution of dark matter
and the relative importance of dark and luminous matter are still questions
under debate. Recent results from the analysis of rotation curves 
(Borriello \& Salucci 2001
for normal spirals,
de Blok et al. 2002 for low surface brightness galaxies) 
seem to favour dark halos with constant density
cores. However the debate is still going on as to whether the available data
have sufficient quality to constrain the distribution of dark matter (Primack 2002).
For this reason high-quality data with high resolution and large spatial 
extension are needed.
This can be ideally achieved by combining
H$\alpha$ rotation curves (for the high resolution) and
HI rotation curves (for the extension to large galactocentric radii).
In particular, the high resolution in the inner parts is crucial to 
deduce the shape of the density profile (Blais-Ouellette et al. 2001),
while the extension to large radii is needed to constrain the size of the 
possible core (Borriello \& Salucci 2001). All these considerations
can be done for the local Universe; the 
Atacama Large Millimeter Array (ALMA) will enable us to investigate the
structure and evolution of dark matter halos at high redshifts.

\section{Observations}

\begin{table} [t]
\begin{center} 
\begin{tabular}{l l l l}
\hline
Galaxy      &  M$_{\rm I}$& R$_{\rm d}$ ('') &  $\theta_{\rm beam}$ ('')\\
\hline
ESO 116-G12 &  -20.0      & 22.7             &  30.4 x 24.4\\
ESO 121-G6  &  -20.4      & 24.4             &  32.2 x 22.7\\
ESO 123-G23 &  -20.8      & 13.2             &  30.6 x 25.0\\
ESO 240-G11 &  -22.3      & 26.3             &  33.8 x 25.6\\
ESO 269-G19 &  -22.3      & 25.4             &  34.3 x 22.0\\
ESO 287-G13 &  -21.7      & 26.3             &  32.6 x 24.6\\
ESO 79-G14  &  -21.4      & 19.0             &  27.4 x 23.0\\
NGC 1090    &  -21.8      & 19.3             &  17.6 x 14.7\\
NGC 7339    &  -20.6      & 17.9             &  13.9 x 13.5\\
\hline
\end{tabular}
\caption{The sample: M$_{\rm I}$ is the absolute magnitude in the I-band,
R$_{\rm d}$ is the exponential scale-length of the stellar disk and
$\theta_{\rm beam}$ is the size of the radio beam}
\end{center}
\end{table}

\vspace{0.3cm}

Our galaxies were selected from the high-quality subsample of H$\alpha$
rotation curves presented in Persic \& Salucci (1995), by finding a
compromise between the following 
criteria: high angular extent
(to maximise $\theta_{\rm beam}/R_{\rm d}$, see Table 1), low luminosity 
(so that the radius where dark matter starts to dominate the 
kinematics is smaller,
see Salucci \& Persic 1999), 
high HI flux (to have better S/N in the HI data) and
symmetry of the H$\alpha$ rotation curve 
(to minimise the effect of non-circular motions).
Our sample currently consists of nine galaxies, and we expect to 
enlarge it in the future.

For the observation and reduction of the optical data (H$\alpha$ 
spectroscopy and I-band photometry) we refer to Persic \&
Salucci (1995) and Mathewson, Ford \& Buchhorn (1992).

The HI observations of NGC 1090 and NGC 7339 were performed
with the VLA in the C-array, while the other galaxies were
observed with the ATCA in the 750m and 1.5km configurations.

\section{Data analysis}

The standard reduction and analysis was performed with the software
packages AIPS, {\it miriad} and GIPSY. 
In Table 1 we list some properties of the galaxies of our sample.
NGC 1090 is the only galaxy of the sample that permits sufficient sampling
of the velocity field to allow the use of the tilted ring modelling 
(Begeman 1989). In all other cases the combination of high inclination
and a large beam (compared with the size of the galaxy) prevents us from 
using this method. We therefore traced the rotation curve 
along the warp on points defined like in Garc\'{\i}a-Ruiz (2001) and
implemented for kinematical use by Vergani et al. (2002).

In all the cases - 
the galaxy suited for tilted ring modelling and the other galaxies -  
we derived the rotation curves by analysing the velocity 
profiles (spectra at single points). In Figure 1 we show an example of a 
typical profile at an intermediate galactocentric distance: the profile
is not symmetrical and it has a tail towards the systemic velocity. This
is what we expect in the case of a highly inclined galaxy,
because (see Fig. 2) what we observe is the integration along 
a large portion of the disk, so that also material with lower radial velocities
contributes to the velocity profile.

Moreover, from recent results (Swaters et al. 1997, Fraternali et al. 2002),
it seems that at least in a few spiral galaxies there is some 
neutral hydrogen at several kpc from the galactic plane that rotates more
slowly than the gas in the disk; this effect produces an asymmetry of 
the velocity profiles (a ``beard'' in the position-velocity diagram,
Fraternali et al. 2002)
also in galaxies that do not have a high inclination.

The result is that the standard methods of first moment (intensity
weighted mean) and single Gaussian fitting provide velocities that 
are biased towards the systemic velocity (see Fig. 1)
and cannot be used to determine the rotation velocity.

\section{The Modified Envelope-Tracing method}

We thus applied a method similar to the Envelope-Tracing
method (e.g. Sofue \& Rubin 2001), 
in order to fit only the side of the profiles that we are
interested in, the extreme velocity side.
We call it {\it Modified Envelope-Tracing} (MET)
method when applied to the whole velocity field and
{\it WArped Modified Envelope-Tracing} (WAMET) method when applied only 
on the ridge of the possible warp: 
\begin{itemize}
\item{we first fitted a half-Gaussian from the peak value 
to the extreme velocity
side of the profiles and we considered the velocity at half maximum ($v_t$)}
\item{then the rotation velocity is given by the following equation:
\begin{equation}
v_{\rm rot}=|v_{\rm t}-v_{\rm sys}|/sin~i-\sqrt{(0.5\cdot {\rm {FWHM}}_{\rm ISM})^2+
({\rm {0.5 \cdot FWHM}}_{\rm instr})^2+
({\rm {0.5 \cdot FWHM}}_{\rm beam})^2}
\end{equation}
where $i$ is the inclination, and the terms indicated as $\rm {FWHM}$
describe the effects that we assume are broadening our profiles:}
\begin{itemize}
\item{ ${\rm {FWHM}}_{\rm ISM}$
is the broadening due to the turbulence of the 
interstellar medium; consistently with Kamphuis (1993) we 
consider a $\sigma_{\rm ISM}$  $={\rm {FWHM}}_{\rm ISM}/\sqrt{8ln2}$
going from 12 km/s in the inner parts to 7 km/s in the outer parts
of the galaxy.} 
\item{ ${\rm {FWHM}}_{\rm instr}$ is the instrumental contribution, 
that we take equal to the channel resolution}
\item{${\rm {FWHM}}_{\rm beam}$ is a rough 
estimate of the broadening of the profiles
due to the beam: as it is evident in Fig. 3, a larger beam will sample a larger portion
of the velocity field and will thus broaden the profiles,
even on their extreme velocity side.\\
According to Sancisi \& Allen (1979) an upper limit to $v_{\rm rot}$ 
can be given 
by setting ${\rm {FWHM}}_{\rm beam}=0$ . \\
To estimate a lower limit to $v_{\rm rot}$, 
in a way similar to Braun (1997) we 
consider that the error due to the beam that we make in measuring the
width of the profiles is given by:\\ 
$2\cdot(v(r \pm  \theta_{\rm beam}/2)-v(r))$ \\
where the positive sign applies when
the gradient of the velocity field is positive and
the negative sign when it is negative. \\
Remembering that we have an upper and a lower limit
to $v_{\rm rot}$, we
decided to take the middle point, i.e. we considered the
following correction: \\
${\rm {FWHM_{\rm beam}}}=v(r \pm \theta_{\rm beam}/2)-v(r)$}

\end{itemize}
\end{itemize}

We preferred to use this method instead of others for numerous reasons:
\begin{itemize}

\item{The profiles are asymmetric, thus methods assuming symmetry cannot be 
used to derive the rotation velocity}
\item{We are interested only in the extreme velocity side of the
profiles, which is not affected by projection effects}
\item{This method accounts for a more realistically varying $\sigma_{\rm ISM}$
with radius,
instead of keeping it fixed}
\item{It also estimates the correction for the broadening
of the profiles due to the beam, which can be substantial in regions 
where the gradient of the velocity field is high}
\item{The WAMET method, applied to galaxies where the sampling of the
velocity field is poor, enables us to trace the rotation curve along the 
possible warp instead of keeping a fixed position angle, like
in methods based on the analysis of the position-velocity diagram; 
the shortcomings of keeping a fixed position angle are discussed in
Vergani et al. (2002)}

\end{itemize}

\section{Rotation curves}
In order to derive the rotation curves, we applied the tilted ring 
modelling to NGC 1090, while for the other galaxies we calculated the kinematical 
centre and the systemic velocity by minimising the differences between the two
sides. The errors are the maximum between three values: the difference
between the approaching and the receding side, our correction for the 
broadening of the profiles due to the beam, and a ``minimum error'' 
equal to ($2/sin~i$) km/s.

In Fig. 3 we show the H$\alpha$ and HI 
rotation curves
projected onto the HI position-velocity diagrams for 4 galaxies
of our sample: 
the agreement between the two datasets
is good and, as we should expect, the HI rotation curves follow the 
shape of the last contours of the
position-velocity diagrams. We also show in Fig. 4 the H$\alpha$ and the 
HI rotation curves derived from three different
methods: our (Warped) Modified Envelope-Tracing method, the first moment
and the single-Gaussian fitting. The two latter methods yield 
a much worse agreement with the H$\alpha$ rotation curves, 
especially in the inner parts where the profiles
are most asymmetric: in these cases the rotation velocities are severely 
underestimated.

From the combined H$\alpha$+HI rotation curves we will
be able to perform the mass decompositions with different models
of the dark halos and to put constraints on the distribution
of dark matter in our sample of spiral galaxies. As a further step,
we plan to supplement and strengthen the method by employing high-resolution 
CO observations.

\vskip 0.8cm

{\noindent
{\bf \large Acknowledgements}

\vskip 0.5cm
GG is grateful for financial support of the 
{\it Deutsche Forschungsgemeinschaft} under number GRK 118 
``The Magellanic System, Galaxy Interaction and the 
Evolution of Dwarf Galaxies''. We thank the
Observatoire de Bordeaux for the warm hospitality
during the workshop.
}

\vskip 0.8cm




{\small
\begin{description}{} \itemsep=0pt \parsep=0pt \parskip=0pt \labelsep=0pt

\item {\bf References}

\vskip 0.5cm

\item Begeman, K.G.: 1989, A\&A 223, 47
\item Blais-Ouellette, S., Amram, P., Carignan, C.: 2001, AJ 121, 1952
\item Borriello, A., Salucci, P.: 2001, MNRAS 323, 285
\item Braun, R.: 1997, ApJ 484, 637
\item de Blok, W.J.G., McGaugh, S.S., Rubin, V.C.: 2001, AJ 122, 2396
\item Fraternali, F., van Moorsel, G., Sancisi, R., Oosterloo, T.: 2002, AJ 123, 3124
\item Garc\'{\i}a-Ruiz, I.: 2001, Ph.D. thesis, Univ, Groningen
\item Kamphuis, J.: 1993, Ph.D. thesis, Univ, Groningen
\item Mathewson, D.S., Ford, V.L., Buchhorn, M.: 1992, ApJS 81, 413  
\item Persic, M., Salucci, P.: 1995, ApJS 99, 501 
\item Primack, J.R.: astro-ph/0205391
\item Salucci, P., Persic, M.: 1999, MNRAS 309, 923
\item Sancisi, R., Allen, R.J.: 1979, A\&A 74, 73
\item Sofue, Y., Rubin, V.: 2001, Ann. Rev. Astron. Astrophys. 39, 137
\item Swaters, R.A., Sancisi, R., van der Hulst J.M.: 1997, ApJ 491, 140
\item Vergani, D., Gentile, G., Dettmar, R.-J., Aronica, G., Klein, U.: 2002, 
astro-ph/0206481, ALMA 
Extragalactic and Cosmology Science Workshop on Dark Matter, Bordeaux, \\
http://www.observ.u-bordeaux.fr/public/alma\_workshop/darkmatter/

\end{description}
}

\begin{figure}[t]
\centerline{{
\vspace {8cm}
\includegraphics{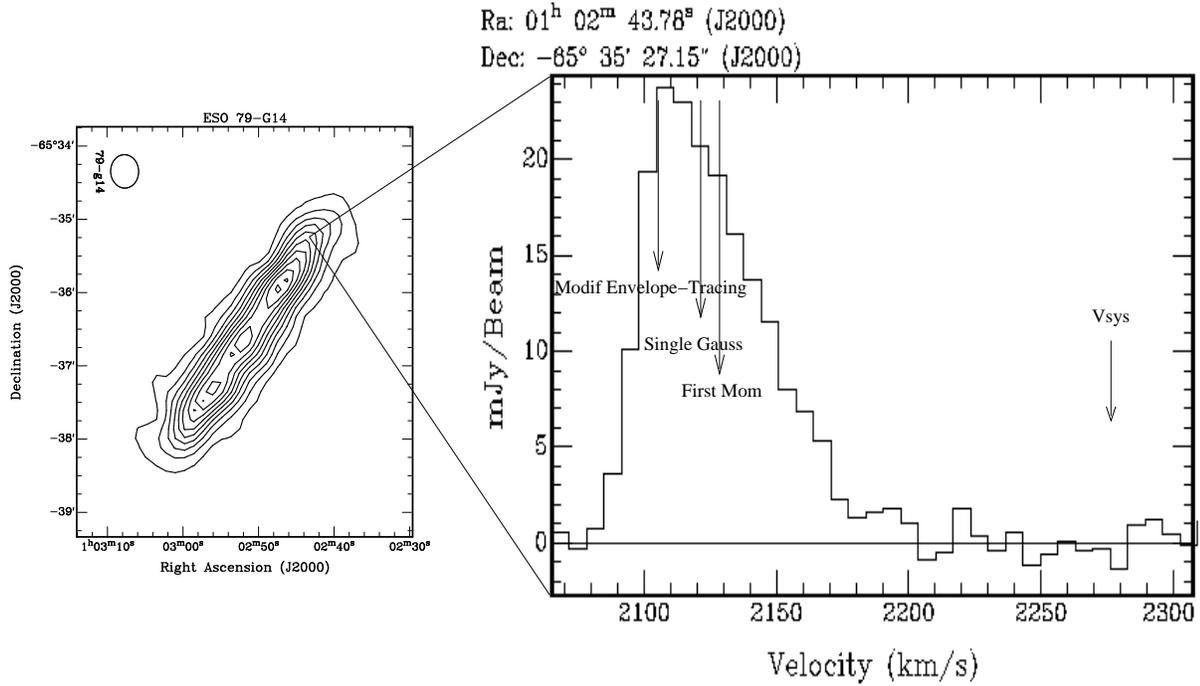}
}}
\caption{A typical spectrum of the galaxy ESO 79-G14 
at an intermediate galactocentric distance.
The arrows indicate the systemic velocity and the positions of the 
velocities derived from the first moment, from fitting
a single Gaussian and from our Modified Envelope-Tracing Method}
\end{figure}

\begin{figure}[b]
\centerline{{
\vspace {4.5cm}
\includegraphics{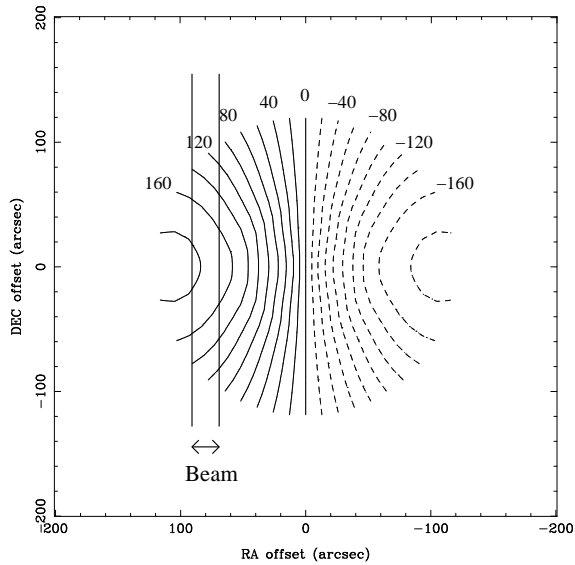}
}}
\caption{Model of the radial velocities in the plane of the galaxy
ESO 79-G14}
\end{figure}

\begin{figure}[t]
\centerline{{
\vspace {16cm}
\includegraphics{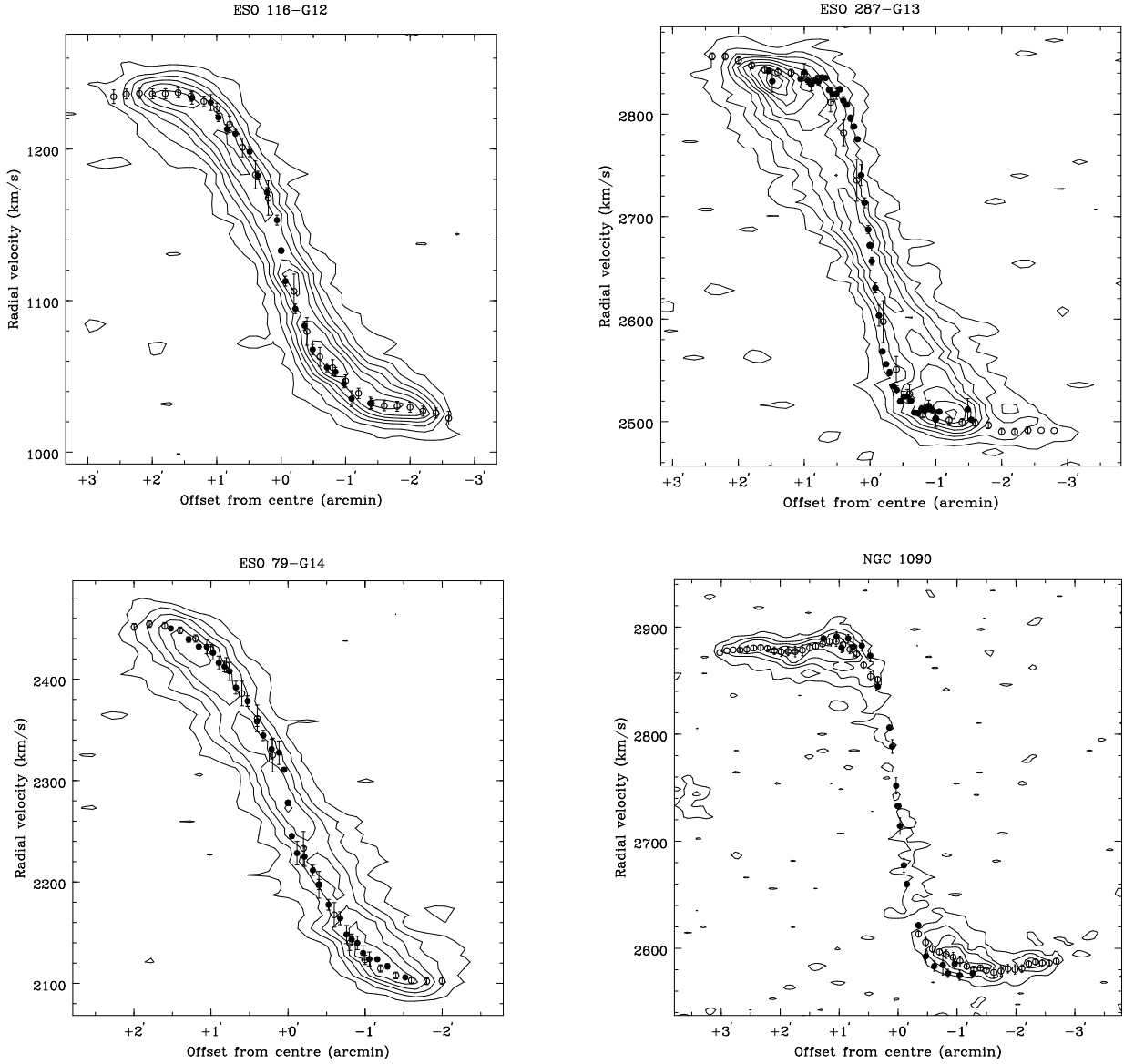}
}}
\caption[]{
H$\alpha$ (filled circles) and HI rotation curves
(empty circles)
projected onto the HI position-velocity diagrams 
of 4 galaxies of our sample. 
We projected sepatately the two sides of the HI rotation curves.
Since the difference between the two sides enters in
out derivation of the error (see text),
at the radii where only one side is present we did not 
define the error.
The contours 
are (2, 6, 10, ...) $\times ~ \sigma$, with $\sigma$=1.2 mJy/beam 
for ESO 116-G12,
ESO 287-G13 and ESO 79-G14 and $\sigma$=0.8 mJy/beam for NGC 1090} 
\end{figure}

\begin{figure}[b]
\centerline{{
\vspace {17cm}
\includegraphics{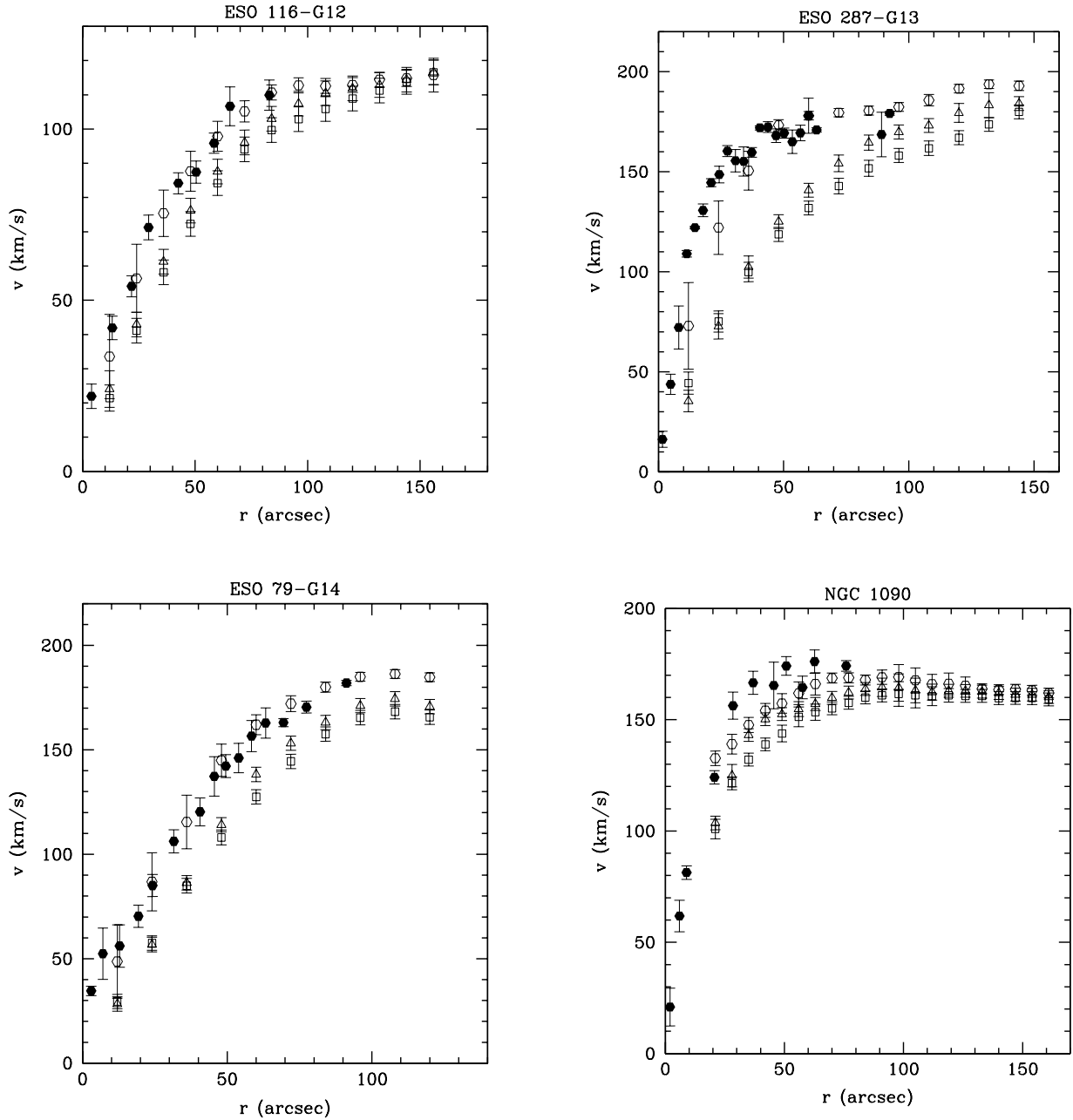}
}}
\caption[]{
H$\alpha$ (filled circles) and HI rotation curves
of the galaxies of Fig. 3 
derived with different methods: our modified method
(empty circles), the first moment (empty squares) and single-Gaussian fitting
(empty triangles)}
\end{figure}

\end{document}